# Structural Origin of Boson Peak in Glasses

Yuan Tian[1]†, Xiaozhe Shen[2]†, Qingyang Gao[1], Zhen Lu[3], Jie Yang[2], Qiang Zheng[2], Christopher Florencio Aleman[1], Duan Luo[2], Alexander Hume Reid[2], Bin Xu[1], Michael Falk[1], Howard Sheng[4], Jianming Cao[5], Xijie Wang[2]*, Mingwei Chen[1]*

[1]Department of Materials Science and Engineering and Hopkins Extreme Materials Institute, Johns Hopkins University, Baltimore, Maryland 21218, United States

[2] SLAC National Accelerator Laboratory, Menlo Park, CA 94025, USA

[3]Advanced Institute for Materials Research, Tohoku University, Sendai 980-8577, Japan

[4] Department of Physics and Astronomy, George Mason University, Fairfax, VA 22030, USA

[5] Physics Department and National High Magnetic Field Laboratory, Florida State University, Tallahassee, Florida 32310, USA

†These authors contributed equally to this paper

*To whom correspondence should be addressed：

E-mails: mwchen@jhu.edu or wangxj@slac.stanford.edu

**Boson peak, the excess low energy excitations in the terahertz regime, is one of the most unique features of disordered systems and has been linked to many anomalous properties of glass materials[1–5]. The nature and structural origin of the boson peak remain elusive and have been debated for more than a half**

**century mainly due to the lack of real-time and real-space experimental insights of the dynamic phenomenon. In this work we employed femtosecond MeV ultrafast electron diffraction[6–8] to characterize the atomic dynamics of metallic glasses in real time. The experiment reveals collective atomic oscillations, presented in elastic electron scattering and atomic pair distribution functions, within the boson peak frequency range of 1.0-1.8 THz in both reciprocal and real space. It was found that the oscillation frequency has reciprocal dependence on interatomic pair distances and the corresponding wave velocity experimentally affirms the transverse acoustic wave nature of the boson peak. The observed strong correlation between THz acoustic vibrations and coherent electron scattering provides compelling evidence that the boson peak originates from the collective transverse vibrational modes of structurally ordered atoms in the disordered system.**

Low energy vibrational excitations are the dynamic origins of thermal and acoustic properties of solid materials, which are generally described by the Debye law with the vibrational density of states $g(\omega) \propto \omega^2$ where ω is frequency. However, it was found that glass materials show anomalous low-temperature thermal behaviors with excess vibrational states, often referred to as the boson peak[1–3], which disobeys Debye's $\omega^2$ law. This unique dynamical property of glasses has been a subject of intense discussion over the past five decades. Many theoretical models have been proposed to explain the nature and origin of the boson peak, such as localized vibrations[5], soft potentials[9–11], elastic heterogeneities[12–15], cohesive cluster resonances[16], phase transitions[17], van Hove singularities[18,19], the Ioffe-Regel limit[3,20,21], and so on. Considering the disordered structures of glasses, these models are usually based on the hypotheses and simulations of structural heterogeneity associated with loosely packed atoms[22,23], low-density defective structures[3,24] or short- and medium-range order[5,16,25]. Though they can each describe certain aspects of the available experimental data, these models are mutually exclusive. As a result, the fundamental question, *i.e.*, the nature and structural origin of the boson peak, remains in dispute up to date.

Experimentally, the boson peak is mainly characterized by inelastic neutron and X-ray scattering[16,18,26–29], Raman scattering[25,30], dielectric spectroscopy[31] and calorimetry[1,23,32]. These experimental approaches have successfully measured the boson peak in momentum and frequency spaces and demonstrated that the boson peak correlates with the chemistries and properties of glasses. For examples, fragility[33], minor doping[22] and plastic deformation[32] can change the intensity and frequency of the boson peak. However, the inherent correlation between glass structure and the boson peak, built on the inelastic scattering and spectroscopic experiments, is far from being determined as direct structural information is absent in these momentum- and frequency-domain experiments. An experiment that is capable of directly linking the fast dynamics and structure of amorphous materials has not previously been realized. Technically, the challenge could be addressed by real-time and real-space characterization of the boson peak. However, this requires both ultrahigh temporal and spatial resolution down to femtoseconds and angstroms, which are not accessible by conventional time-resolved diffraction and imaging techniques. In this study we employed the state-of-the-art MeV ultrafast electron diffraction (MeV-UED) technique[7,34] to study the fast dynamics of metallic glasses (MGs). By using femtosecond (fs) optical pumping to excite the resonant modes of the boson peak, we observed the collective vibrational modes of the boson peak in the time-domain by coherent MeV electron diffraction and corresponding atomic pair distribution function (PDF) analysis, unveiling the nature and structural origin of the excess low energy excitations in metallic glasses.

The MeV-UED experimental setup is schematically drawn in **Figure 1(a)**. $Zr_{55}Cu_{30}Ni_5Al_{10}$ MG thin films were pumped by fs laser pulses and the induced structural dynamics were detected by a MeV electron probe. The total electron scattering intensity was recorded as a function of pump-probe delay time. A low pump fluence of ~5 mJ/cm$^2$ was used to avoid possible structural damage of the glass samples. The temperature rise induced by the ~5 mJ/cm$^2$ laser fluence is estimated to be ~8.3K (*see* **Supplementary Information**) and the sample temperature is far below the glass transition point (~690 K) of the alloy[35] during the MeV-UED experiments. Consistent with conventional electron diffraction patterns of MGs, the time-resolved MeV-UED patterns (**Fig.1(a)**) show halo rings lying symmetrically over the central beam spot. The azimuthal integration was employed to convert two-dimensional MeV-UED patterns to one-dimensional diffraction intensity curves for quantitative studies of the glass structure at different time delays (**Fig. 1(b)**). The magnified profiles of the first diffraction peak (the inset of **Fig. 1(b)**) display visible peak shift and maximum intensity decrease with time delays, which are in accordance with the typical thermalization process of MGs and crystalline metallic materials stimulated by fast laser pumping[36]. The temperature rise induced by the laser stimulation initiates intensified atomic vibrations and thereby the decrease of the MeV-UED peak intensity due to the Debye-Waller effect. Meanwhile, the temperature rise also leads to thermal expansion, *i.e.,* increase of equilibrium interatomic distances, as reflected by the first peak shift to lower $Q$[37].

The entire dataset of the time-dependent diffraction profiles is shown in a false-color plot in **Supplementary Figure 1**. The diffraction peaks can be well fitted by the Lorentzian function (**Fig. 2(a)** and **Supplementary Figure 2**). This allows us to conduct quantitative analysis on structural dynamics by fitting the diffraction peaks at different time delays (*see* **Supplementary Information**). For the first diffraction peak, the time-dependent peak area, maximum peak intensity, full width at half maximum (FWHM), and peak center position are depicted in **Fig.2(b-d)** respectively, together with exponential decay fittings based on the two-temperature model[38] and Debye-Waller effect (*see* **Supplementary Information**). The time dependences of the three diffraction parameters all show the typical tendency of thermalization. The gradual decay of peak area (**Fig.2(b)**) suggests enhanced random thermal scattering by intensified atomic vibrations, *i.e.,* excitations of thermal phonon modes, from the Debye-Waller effect. The baselines can be well fitted by an exponential function. The absence of any further change of peak area after ~1 picosecond (ps) indicates the end of the thermalization process. In contrast to the negligible peak width changes observed in crystals, the broadening of the FWHM (**Fig.2(c)**) in the glass may be associated with non-affine atomic displacements in the disordered system during thermalization[39]. The shift of the diffraction peak position with the time delays results from thermal expansion with increased average interatomic distances (**Fig.2(d)**). Based on the time-dependent peak shift, the volume expansion rate is estimated to be $3.62 \times 10^{-4}$/ps, suggesting that the overall sample volume change is negligible during MeV-UED experiments with the maximum delay time of 1.73 ps. From the exponential decay fittings of the peak area, the characteristic relaxation time for the thermalization was determined to be ~0.4ps (*see* **Supplementary Information**). This is significantly faster than that of crystalline materials due to the relaxation of momentum conservation for electron-phonon scattering in the disordered material[36,40]. Therefore, the "phonons" with large density of states (DOS) in the glass play the dominant role in electron-phonon thermalization.

Interestingly, unlike the monotonic decay of the diffraction peak area and intensity during the thermalization process of crystalline metallic materials[41], intensive collective oscillations, superimposed on the exponential decay fitting curves of the peak intensity, FWHM and positions, can be observed (**Fig.2(c, d)**). The THz oscillations become more obvious after subtracting the monotonic exponential decay from the original time-dependent diffraction intensity curves (**Supplementary Figure 3 and 6**). The nearly identical oscillation frequencies (1.5-1.6 THz) of the peak intensity, FWHM and position suggest the same physical origin leading to the collective atomic vibrations in the disordered system. Carefully analyzing the intensity changes with time delays, we noticed that the first diffraction peak center at $Q$=2.7Å$^{-1}$ and the left and right wings at $Q$=2.0Å$^{-1}$ and 3.4Å$^{-1}$ show the strongest intensity oscillations with the frequency of ~1.5-1.8 THz (**Fig.2 (e-g)**). The intensity oscillations can be well fitted by the sine function as plotted in solid lines, suggesting a sinusoidal waveform of the collective oscillations in the disordered material. The intensity oscillatory behaviors of the two wings have an approximate 180° (π) phase lag with that of the peak center (**Fig.2 (e-g)**), which is in line with the phase relation between peak intensity and FWHM in **Fig.2 (c)**. Thus, the first diffraction peak undergoes sinusoidal narrowing and broadening oscillations akin to a normal mode of an oscillating system. The observed frequency of ~1.5-1.8 THz falls in the range of the excess vibrational mode of the boson peak of Zr-based MGs (*see* **Supplementary Table 1**). As no any other coherent vibration and relaxation modes have been reported in this frequency range for glasses, we anticipate that the observed collective atomic oscillations are the resonant modes of the boson peak[10,11], which are stimulated by the femtosecond optical pulse and that overpower other vibration modes as the dominant one in the MeV-UED experiments. The assumption that the collective atomic oscillations correspond to the resonant boson peak modes is also supported by molecular dynamics simulations that exhibit resonance in the boson peak frequency regime during cyclic straining of a metallic glass sample (**Supplementary Figure 9**). Note that the observed THz oscillations, arising from the intrinsic vibrational modes of metallic glasses, are fundamentally different from the volumetric acoustic waves generated by short stress pulses from fs laser, which have frequencies in the gigahertz regime for films tens of nanometer thick (**Supplementary Figure 8**)$^{36,42}$.

In addition to the first diffraction peak, the intensity oscillations can also be observed from high-order elastic diffraction peaks. We found that the intensity oscillatory behaviors appear at multiple *Q* values, which always locate at the centers and wings of the high-order diffraction peaks as marked in **Fig.3(a)**. All the high-*Q* intensity oscillations have the sinusoidal waveform and can be well fitted by the sine function in dashed lines. Again, the collective intensity oscillations of the peak centers and wings keep the approximate anti-phase relation as those of the first peak (**Fig.3(b)**). We converted the real-time intensity oscillations at different *Q* to the frequency domain by Fourier transformation (**Supplementary Figure 7(a)**), which is well consistent with the boson peak modes of Zr-based metallic glasses measured by inelastic neutron scattering (**Supplementary Figure 7(b)).** This further verifies that these collective oscillations detected by MeV-UED are the resonant modes of the boson peak. The strong coupling between the collective oscillations and the coherent electron scattering provides compelling evidence that the boson peak is closely related to local structure order of the metallic glass. The cooperatively arranged atoms in the disordered system have in-phase relations from local rotation and/or translation symmetries for the collective vibrations and the constructive interference of electron scattering[43]. Since the fixed phase relations cannot be reserved in propagative modes that only lead to monotonic changes in diffraction peak width (*see* **Supplementary Information**), the well-defined phase relation of the oscillations between centers and wings within each diffraction peak also demonstrates that the observed vibrational modes are non-propagative and are strongly confined to local structures formed by atoms with the in-phase relations presented by the coherent electron diffraction. In fact, the localized nature of the observed vibrational modes is in line with the assumption that the observed collective atomic oscillations are the resonant modes of the boson peak as the localized standing waves are usually associated with resonance at an eigenfrequency. We noticed that the frequency of the collective oscillations from the left wing to the peak center and the right wing within the 1$^{st}$ diffraction peak changes from 1.5 THz to 1.8 THz with $Q$ from ~2.0 Å$^{-1}$ to ~4.0 Å$^{-1}$ in a nearly linear manner (**Supplementary Figure 4**). Beyond the first diffraction peak ($Q$>4.0Å$^{-1}$), the oscillation frequency varies in a wider range from 1.2 THz to 1.8 THz and does not have a well-defined $Q$ dependence. It appears that the $Q$ dependence is only preserved within each diffraction peak regime.

Comparing with inelastic X-ray and neutron scattering and spectroscopies that only display the boson peak in the momentum and frequency domains, the time-resolved elastic electron scattering of MeV-UED can be analyzed by the difference atomic pair distribution function (ΔPDFs) method considering both Bragg diffraction and diffuse scattering[8,44] (details in **Supplementary Information**) to illustrate the THz atomic dynamics of the MG in real space. Using the time-zero PDF profile, *i.e.*, right before laser pumping, as the reference, we plotted the 3D diagram of time-dependent ΔPDF profiles to show the amplitude decrease (plotted as positive) and increase (as negative) with time delays up to the 4$^{th}$ atomic pair distance $r$ =~1.0 nm (**Fig.4(a)**). The red curve in **Fig.4(a)** is the static PDF profile of the $Zr_{55}Cu_{30}Ni_5Al_{10}$ MG derived from synchrotron X-ray diffraction (XRD) measurements, which gives the guidance for determining the atomic pair radii in the ΔPDFs. The obvious changes of ΔPDF amplitudes with the time delays are all within the peak regions of the red PDF profile, indicating amplified atomic vibrations during thermalization. Consistent with the intensity oscillations of diffraction peaks in reciprocal space, the periodic amplitude oscillations are observed from the 1$^{st}$ to 4$^{th}$ atomic pair peaks in the 3D ΔPDF diagram (**Fig.4(a)**). Particularly, more obvious oscillations can be observed in the 2$^{nd}$ atomic pair region with $r$ from ~4.0 Å to ~6.5 Å. The oscillation signals at 1$^{st}$ to 4$^{th}$ atomic pair distances are extracted and plotted in **Fig.4 (b)** by removing exponential decay backgrounds using the method described in the **Supplementary Figure 3**. The ΔPDF amplitude oscillations also exhibit a sinusoidal waveform with frequencies of 1.0-1.8THz, coinciding with the observations in reciprocal space. In the 1$^{st}$ atomic-pair ΔPDF peak region that corresponds to the density distribution changes of the nearest neighboring atoms, the oscillations can be well identified from the left and right wings at $r$=2.3 Å and 4.0 Å with a nearly constant frequency of ~1.7 THz. However, an obvious oscillation cannot be observed at the maximum of the peak, indicating that the vibrational coherence may be less prominent at the nearest neighbor length scale and incoherent thermal vibrations (Debye-Waller effect) may dominate the peak intensity variation. On the other hand, the oscillations at the left and right wings are mainly from peak broadening which is less influenced by the Debye-Waller effect.

The real-space and real-time ΔPDFs enable us to directly measure the dependence of the oscillation frequency on the correlated atomic-pair distance $r$. The frequency gradually decreases from ~1.8 THz to ~1.0 THz from $r=$ ~4.0 Å to ~10.0 Å, which can be well fitted by the reciprocal function: $f = v/2r$, where $f$ is the oscillation frequency and $v$ is a constant (**Fig.4(c)**). The coupling between the wavelength $v/f$ and the correlated interatomic distance $r$, verified by MD simulation (**Supplementary Figure 9**), yields a value of $v$ equal to 1747±69 m/s. This is very close to the reported transverse acoustic wave velocity in Zr-based MGs (~1750 m/s)[45], suggesting the transverse acoustic wave nature of the boson peak in the MG. The real-time real-space experimental result is consistent with previous computational simulations of hard sphere glasses and MGs that the boson peak correlates with transverse acoustic waves[3] as the boson peak frequency coincides with the Ioffe-Regel limit of transverse phonons[3,46,47]. The reciprocal dependence and the transverse acoustic wave nature further demonstrate that these oscillation modes are stationary waves and affiliated with the local structures of the glass to which they are confined[5]. It is interesting to note that the oscillation frequencies within the 1st atomic-pair peak region ($r<$ ~4.0 Å) deviate from the reciprocal relationship and appear to be independent of the correlated interatomic distance $r$. The occurrence of the $r$ dependence starting from the 2nd atomic-pair peaks indicates the strong scattering and localization of transverse acoustic phonons in the 2rd atomic pair region ($r=$ ~4.0 Å to ~6.5 Å)[3,20,21,46] and the collective oscillations of the nearest neighboring atoms in 1st PDF peak are dominated by the stationary wave localized in the second nearest atomic neighbors.

As a unique dynamic behavior of disordered systems, the boson peak has been widely attributed to local defective or ordered structures in disordered glasses[3,5,16,25]. These local structural features unsurprisingly lead to nanoscale heterogeneity in density and elasticity[48,49], which have been the foundation of many boson peak models as they can lead to phonon localization, Ioffe–Regel crossover, heterogeneous elasticity, and so on[3,5,13,16]. However, the direct correlation between these local structures and the boson peak has not been illustrated by an experiment. Our MeV-UED experiments reveal the THz collective atomic oscillations in the boson peak frequency range in both reciprocal and real spaces. Importantly, the collective oscillations are measured with coherent elastic electron scattering and at the correlated atomic positions in PDF profiles. Thus, the THz atomic vibrational modes are associated with collective motions of atoms that have in-phase cooperative relations in the disordered material as they can generate the constructive interference for electron scattering. From the real-space and real-time ΔPDF analyses, we observed that the collective vibrations are conspicuous on the length scale ($2r$ = 0.8-1.3 nm) of the 2$^{nd}$ atomic-pair distance, which is the typical length scale of medium-range order (MRO) in amorphous materials[5,25,50]. Combining with the reciprocal dependence of oscillation frequency on the correlation length of local cooperative structures, we can conclude that the boson peak vibrational modes are stationary waves localized to the MRO structures, which is in line with the transverse acoustic nature. Moreover, with the collective oscillation measurements in both real and reciprocal spaces, we can gain the insights on the dynamic correspondence of the diffraction peaks (especially the first strongest diffraction peak) and real-space atomic structure. In reciprocal space, the first diffraction peak has an intensity oscillation frequency ranging from 1.5-1.8THz. Harmoniously, this frequency range corresponds to the $2^{nd}$ atomic pair distance peak in the PDF. Therefore, the dynamic correspondence indicates that the first diffraction peak is mainly contributed by the constructive interference from elastically scattered electrons from the cooperatively arranged atoms with the coherent length of about the second nearest neighbor distance. For MGs that are mainly constituted by transition metal atoms, the length scale is generally ~0.8-1.5 nm in diameter. Moreover, the fact that the identical collective vibration modes can be observed in both real and reciprocal spaces indicates that Bloch's theorem still hold in the disordered system for the collective effect of multiple atoms that have local in-phase translational invariance. Therefore, based on the real-time MeV-UED experiments, we could conclude that the boson peak originates from the cooperatively arranged atoms that have the in-phase relations in elastic diffraction in a medium-range correlation length, and the localized states of the boson peak couple predominantly to the transverse acoustic modes.

**Methods**

**Sample preparation and TEM characterization**

A $Zr_{55}Cu_{30}Ni_5Al_{10}$ alloy with excellent glass forming ability is chosen for the MeV-UED experiments. The $Zr_{55}Cu_{30}Ni_5Al_{10}$ glass films with a thickness of ~35nm were prepared by single-target radio frequency magnetic sputtering at a power of 50W. The working argon pressure was set at 0.3Pa and the substrate was at room temperature. The glass nature of the as-sputtered samples was confirmed using a FEI TF30 transmission electron microscopy (TEM) operating at 300kV. TEM characterization depicted in **Supplementary Fig.5(a, b)** verified that the samples are fully amorphous. The glass transition temperature ($T_g$) of the $Zr_{55}Cu_{30}Ni_5Al_{10}$ glass films is ~690K[35].

**MeV-UED experiments**

The MeV-UED experiments were conducted at SLAC National Accelerator Laboratory. The details on the MeV-UED setup can be found in ref. 7 and 34. In this experiment, the probing electron energy was 3.7 MeV and the electron scattering patterns were recorded in the range of $1.0 < Q < 11.7$ Å$^{-1}$ by an electron detector. Pair distribution functions were employed to obtain real-space structural information. The high-energy MeV electron diffraction has obvious advantages in high $Q$ and $r$ resolutions, a relatively flat Ewald sphere and less multiple-scattering for total electron scattering analysis. Laser pulses with a wavelength of 400 nm and a pulse duration of 60 fs (FWHM) were used for stimulating the glass samples and for generating electron probes. The MeV-UED facility has the temporal resolution of ~70 fs [root mean square (rms)] and reciprocal space resolution of 0.039 Å$^{-1}$ (rms)[7]. The high spatial and temporal resolutions of the MeV-UED system have been well demonstrated in previous studies[8]. Time-dependent diffraction was measured in a series of time delays from -0.3 ps to +1.73 ps with a scan step of 67 fs. Multiple scans were conducted over the entire delay time range. The experiments were conducted at a frequency of 120Hz and at a constant environmental temperature of 300±1 K. In each scan, diffraction patterns at each time delay were accumulated for 10 seconds. In this study we used a low laser fluence of 5 mJ/cm$^2$ to avoid any irreversible structural relaxation. A momentum-space Gaussian filter (0.055 Å$^{-1}$ FWHM) plus a temporal Gaussian filter (160fs FWHM) were applied on the dataset to suppress noise and highlight characteristic features in structure and dynamics.

**Data Availability**

Experimental data were generated at the MeV-UED facility at the SLAC National Accelerator Laboratory. Data behind each figure are available at https://engineering.jhu.edu/mingweichen/publications/. Raw datasets are available from the corresponding authors on request.

**Code availability**

The non-commercial codes used for the simulation and analysis here are available at https://engineering.jhu.edu/mingweichen/publications/.

**Acknowledgments** We thank Prof. Takeshi Egami, Prof. Todd Hufnagel, Prof. Ryan Hurley, Prof. Jim Spicer, and Prof. Kevin Hemker for valuable discussion. This work was supported by the National Science Foundation (NSF DMR-1804320 and DMR-1910066/1909733) with Dr. Judith Yang as the program director, and the U.S. Department of Energy contract no. DEAC02-76SF00515. The experiment is performed at SLAC MeV-UED and supported in part by the U.S. Department of Energy (DOE) Office of Science, Office of Basic Energy Sciences, SUF Division Accelerator & Detector R&D program, the LCLS Facility, and SLAC under contract Nos. DE-AC02-05CH11231 and DE-AC02-76SF00515. The MD simulations were carried out at the Advanced Research Computing at Hopkins (ARCH) core facility (rockfish.jhu.edu), which is supported by the National Science Foundation (NSF) grant number OAC 1920103.

**Author contributions** M.C. conceived and supervised the research. X.W. supervised MeV-UED experiments. Y.T., X.S., Q.Z., D.L. and A.H.R. performed MeV-UED experiments. Y.T., B.X., H.S. and M.F. contributed to MD simulations. Z.L. prepared the metallic glass thin film samples. J.C., X.S., Q.G., J.Y., C.F.A., H.S. and Q.Z. contributed to the data analysis. M.C. and Y.T. conducted data analysis and wrote the manuscript. All the authors contributed to result discussion and manuscript writing.

**Correspondence and requests for materials** should be addressed to M.C. or X.W.

**Competing interests** The authors declare that they have no competing interests.

**Supplementary information** This file contains Supplementary Methods, Supplementary Notes 1-4, Supplementary Figure 1-9, Supplementary Table 1 and additional references. Supplementary Notes 1-5 present details of: i) Estimation on temperature rise during laser pumping; ii) Calculations on the relaxation time of thermalization; iii) Analysis on the localized nature of the observed oscillation mode; and iv) Molecular dynamics simulations of dynamic mechanical spectroscopy.

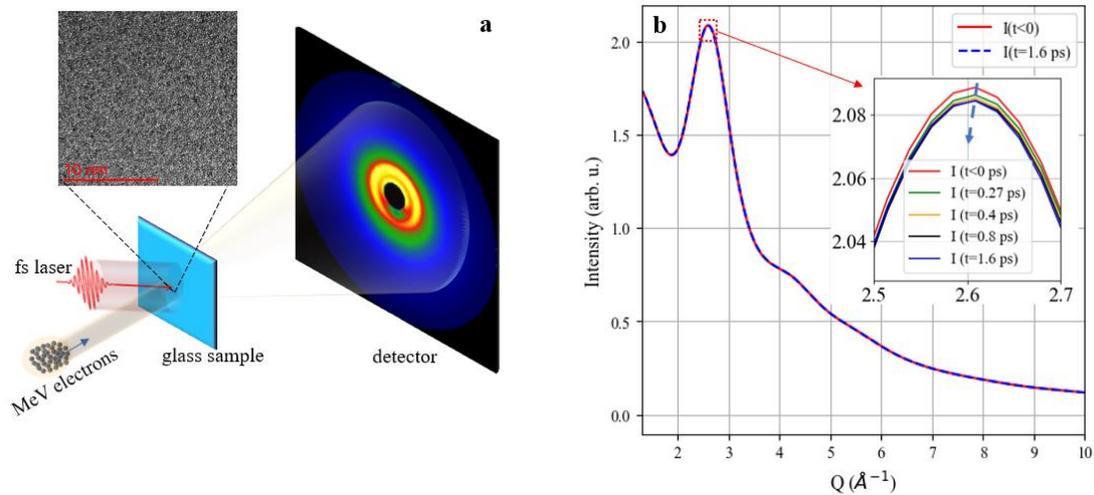

**Figure 1. Schematics of the MeV-UED experiment setup and data collection.** (a) Schematic drawing of the experimental setup. (b) The electron diffraction profiles at time zero (-0.3 ps before laser pump) and 1.5 ps. The inset image shows the evolution of the first diffraction peak with time delays.

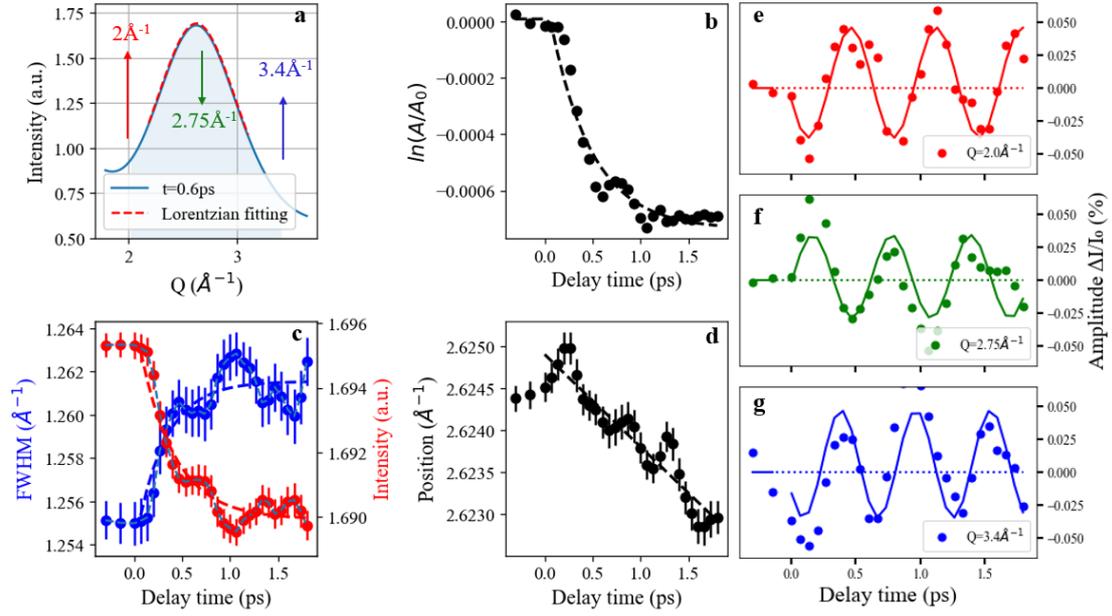

**Figure 2. Temporal evolution of the first diffraction peak.** (a) Lorentzian fitting curve on first diffraction peak. (b) Peak area evolution in the $Q$ range between 2 Å$^{-1}$ and 3.4 Å$^{-1}$ over delay time. The time dependence of the logarithm of peak area is fitted to exponential decay function, $\ln(A/A_0)=a[1-\exp(-t/t_0)]+b$, as plotted in dashed line. (c) FWHM and peak intensity evolutions over delay time with exponential decay fitting curves in dashed lines. (d) Peak position shift over delay time with exponential decay fitting curve in dashed line using equation $Q_{max}=a[1-\exp(-t/t_0)]+b$. Error bars correspond to one standard deviation of the Lorentzian fitting. (e-g) Diffraction intensity oscillation over delay time at $Q=2.0$ Å$^{-1}$, 2.75 Å$^{-1}$ and $Q=3.4$ Å$^{-1}$. Solid lines represent the sine fitting of the corresponding data points with equation $\Delta I(q)/I_0(q)=a\sin(\omega t+\phi)$.

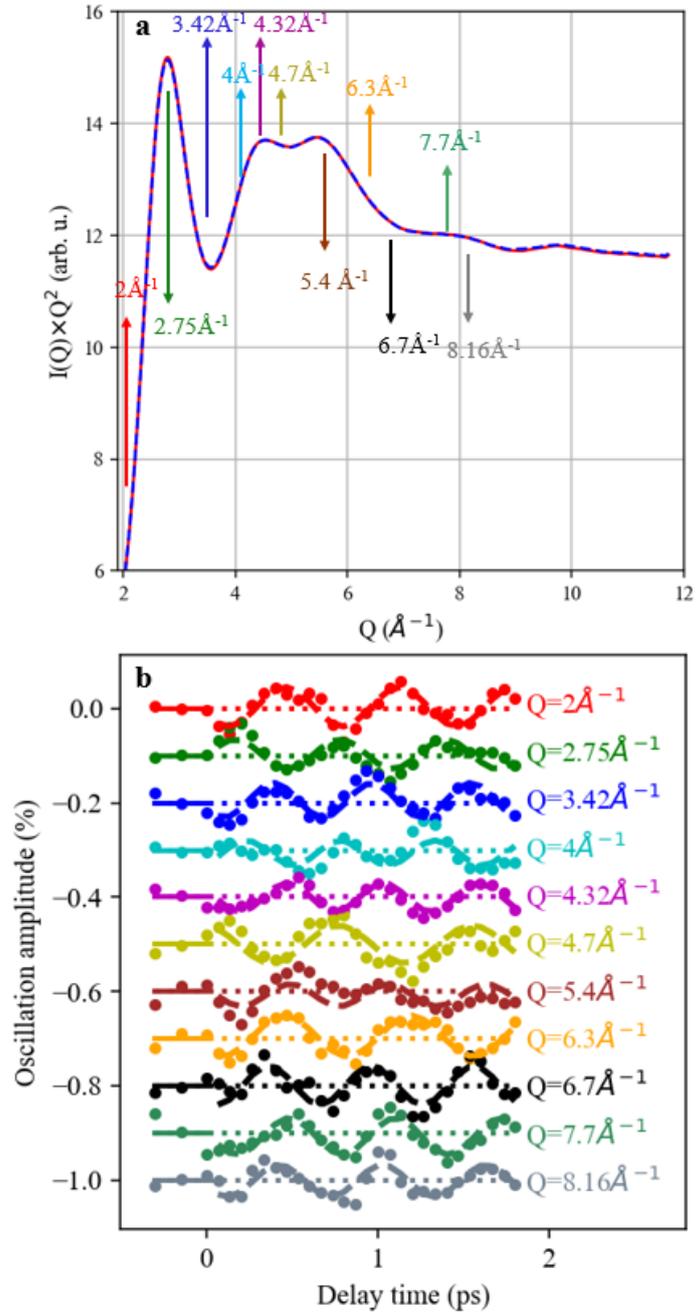

**Figure 3. Collective intensity oscillations of MeV-UED diffraction peaks of Zr$_{55}$Cu$_{30}$Ni$_5$Al$_{10}$ MG.** (a) $I(Q) \cdot Q^2$ plot of the diffraction profile for highlighting peaks at high $Q$ values. (b) Diffraction intensity oscillations with delay time at different $Q$. Dashed lines represent the sine fitting with the equation $\Delta I(q)/I_0(q) = a\sin(\omega t + \phi)$.

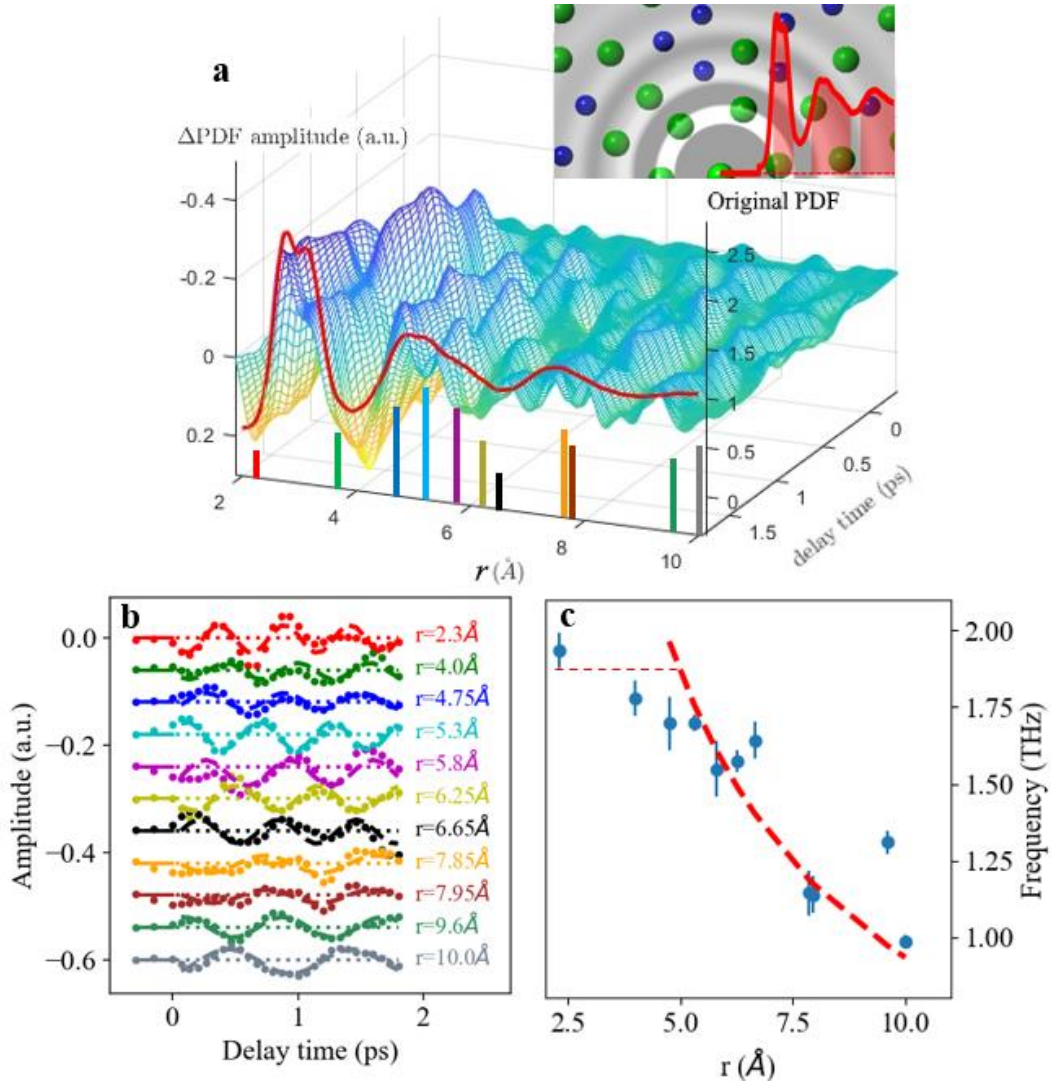

**Figure 4. Collective intensity oscillations in time-resolved ΔPDF profiles of $Zr_{55}Cu_{30}Ni_5Al_{10}$ MG.** (a) Three-dimensional diagram of time-dependent ΔPDF profiles. The red curve is the PDF profile of the $Zr_{55}Cu_{30}Ni_5Al_{10}$ MG derived from synchrotron XRD. The inset image on the top right corner is the schematics of PDF which represents the average density of neighboring atoms at corresponding interatomic distances. (b) Intensity oscillations at different *r* positions marked in (a). The Sine function fittings on the collective oscillations are plotted in dashed lines with the equation: $\Delta PDF = a\sin(\omega t + \phi)$. (c) The frequency of the sine oscillations at different *r* positions. The red dashed line shows the reciprocal function fitting on the $\omega$-r relation using the method of least squares. The oscillations near the first peak are excluded for the non-recognizable oscillations in the center regime. Error bars correspond to the standard deviations of the sine function fittings in (b).